\documentclass[11pt,twoside]{article}

\usepackage{ihep,epsfig,amsfonts,amssymb,amsbsy}

\newcommand{\be}{\begin{equation}}
\newcommand{\ee}{\end{equation}}
\newcommand{\beas}{\begin{eqnarray*}}
\newcommand{\eeas}{\end{eqnarray*}}
\newcommand{\bea}{\begin{eqnarray}}
\newcommand{\eea}{\end{eqnarray}}
\newcommand{\ba}{\begin{array}}
\newcommand{\ea}{\end{array}}

\newcommand{\al}{\alpha}

\newcommand{\g}{\gamma}
\newcommand{\de}{\delta}

\begin{document}

\title{\bf \uppercase{Space-time: emerging vs.\ existing}
}
\author{ \bf {Yu.\ F.\ Pirogov}\\
{\it  Institute for High Energy Physics, }
{\it  Protvino,  RU-142281, Russia}
}
\date{}
\maketitle

\begin{abstract}
The concept of the space-time as emerging in the world
phase transition, vs.\ a~priori existing, is put forward. 
The theory of gravity with two basic symmetries, the
global affine one and the general covariance, is developed.
Implications for the Universe are indicated.
\end{abstract}

\section*{Introduction}

Conventionally, the physical sciences start with the  space-time
equipped with  the metric as the inborn structure. It is proposed in
what follows to substitute the metric space-time by  the world
continuum which possesses only the affine connection.
The metric is to emerge spontaneously at the effective level during
the world structure formation. Ultimately, this  results in 
the theory of gravity, the Metagravitation, based on two basic
symmetries -- the global affine one and the general covariance, with
the graviton being the tensor Goldstone boson. For more detail, see
ref.~\cite{Pirogov}. This is the physics implementation of
the approach to gravity as the nonlinear model
$GL(4,R)/SO(1,3)$~\cite{Salam,Ogievetsky}.
Generically, the concept developed is much in
spirit of the ideas due to E.~Schr\"odinger~\cite{Schrodinger} and
I.~Prigogin~\cite{Prigogin}.

\section*{Affine symmetry}

\paragraph{Affine connection}

Postulate that the predecessor of the space-time
is the world continuum equipped only with the affine
connection. Let~$x^\mu$, $\mu=0,\dots,3$  be the world coordinates.
In ignorance of the underlying theory,
consider all the structures related to the
continuum as the background ones. Let
$\bar\psi^\lambda{}_{\mu\nu}(x)$ be the  background
affine connection. Let the antisymmetric part of the connection   be
absent identically.
Let $P$ be a fixed but otherwise arbitrary point  with the world
coordinates~$X^\mu$. One can annihilate the connection in this point
by adjusting the proper coordinates
$\bar\xi^{\al}(x,X)$. In the vicinity of~$P$, the connection now
becomes
\be\label{eq:connection}
{\bar \psi}^{\g}{}_{\al\beta}(\bar\xi)= \frac{1}{2}\,
{\bar \rho}^{\g}{}_{\al\delta\beta}(\bar\Xi)\,
(\bar\xi-\bar\Xi)^{\de} + {\cal O}((\bar\xi-\bar\Xi)^2),
\ee
with ${\bar \rho}^{\g}{}_{\al\delta\beta}(\bar\Xi)$
being the background  curvature tensor in the 
point~$P$ and $\bar\Xi^{\al}= \bar\xi^{\al}(X,X)$.
Let us consider the whole set of the  coordinates
with the proper\-ty $\bar \psi^{\al}{}_{\beta\g}\vert_P=0$.
The allowed group of transformations of such 
coordinates is the inhomogeneous general linear group
$IGL(4,R)$ (the affine one):
\be\label{eq:affine}
(A,a)\ : \ \bar\xi^{\al} \to
\bar\xi'^{\al}=A^{\al}{}_{\beta}
\bar\xi^{\beta}+a^{\al},
\ee
with $A$ being an arbitrary nondegenerate matrix.
Under these, and only under these transformations, the affine
connection in the point~$P$ remains to be zero. The group is
the global one in the sense that it transforms  the
$P$-related coordinates in the global manner,  i.e., for all the
continuum at once. The respective coordinates will be called the local
affine ones. In these
coordinates, the continu\-um in a neighbourhood of the  point
is approximated by  the affinely flat manifold. In particular,  the
underlying covariant derivative in the affine coordinates in the
point~$P$ coincides with the partial derivative.

\paragraph{Metarelativity}

According to the special relativi\-ty, the present-day physical laws
are invariant relative to the choice of the inertial coordinates, with
the symmetry group being the Poincare one.
Postulate the principle of  the extended relativity, the
Metarelativity, stating
the invariance  relative to the choice of the affine coordinates.
The  symmetry group extends now to the affine one. The
latter is 20-parametric and expand  the 10-parameter Poincare
group $ISO(1,3)$ by the ten special affine transformations.
There being no exact affine symmetry, the latter should be broken to
the Poincare symmetry in transition from the underlying  to 
effective level.

\paragraph{Metric}

Assume that the affine symmetry breaking is achieved due to the
spontaneous emergence  of the background
metric $\bar\varphi_{\mu\nu}(x)$ in the world
continuum. The metric is assumed to have the Minkowskian signature and
to look like
\be\label{eq:metric}
\bar \varphi_{\al\beta}(\bar\xi)=
\bar\eta_{\al\beta}-\frac{1}{2}
\bar
\rho_{\g\al\delta\beta}(\bar\Xi)\,
(\bar\xi-\bar\Xi)^{\g}(\bar\xi-\bar\Xi)^{\de}+{\cal
O}((\bar\xi-\bar\Xi)^3).
\ee
Here one puts   $\bar\eta_{\al\beta}\equiv \bar
\varphi_{\al\beta}(\bar\Xi)$ and 
$\bar \rho_{\g\al\delta\beta}(\bar\Xi)=
\bar \eta_{\g\de}\,
\bar \rho^{\de}{}_{\al\delta\beta}(\bar\Xi)$.
The metric~(\ref{eq:metric}) is such that the Christoffel
connection $\bar\chi^\g{}_{\al\beta}(\varphi)$, determined by the
metric, matches with the  affine
connection $\bar\psi^\g{}_{\al\beta}$ in the sense that the
connections coincide locally, up to the first derivative:
$\bar\chi^\g{}_{\al\beta}=\bar\psi^\g{}_{\al\beta}+{\cal
O}((\bar\xi-\bar\Xi)^2)$. 
This is reminiscent of the fact that the metric in
the Riemannian manifold may be  locally  approximated, up to the
first derivative, by the flat metric. Associated  with the
background  metric, there appears the partition of the amorphous
4-dimensional continuum onto the space and time. 
 
Under the  affine symmetry, the  background metric
ceases to be invariant. But it still possesses an
invariance subgroup. Namely, without any loss of
generality, one can choose among the  affine
coordinates the particular ones with $\bar \eta_{\al\beta}$
being in the Minkowski form $\eta=\mbox{diag}\,(1,-1,-1,-1)$. The
respective coordinates will be called the background inertial
ones. They are to be distinguished from the effective inertial
coordinates (see later on). Under the affine transformations, one has 
\be\label{eq:eta_lin}
(A,a)\ : \ \eta\to \eta'=A^{-1T}  \eta A^{-1}\ne\eta, 
\ee
whereas  the Lorentz transformations $A=\Lambda$ still leave $\eta$
invariant.
It follows that the subgroup of invariance of $\eta$ is the Poincare
group $ISO(1,3)\in IGL(4,R)$. 
Under the appearance of the metric, the $GL(4,R)$ group is
broken spontaneously to  the residual Lorentz one
\be
GL(4,R)\stackrel{M_A\ }{\longrightarrow} SO(1,3),
\ee 
with the translation subgroup being intact.
For the symmetry breaking scale $M_A$, one expects a priori
$M_A\sim M_{Pl}$,  with $M_{Pl}$ being the Planck
mass.

\paragraph{Affine Goldstone boson}

Attach to the  point~$P$  
the auxiliary linear space~$T$, the tangent space in the point.
By definition, $T$ is isomorphous to the Minkowski
space-time. The tangent space is the
structure space of the theory, whereupon the
realizations of the physics space-time
symmetries, the affine and the Poincare ones, are implemented. 
Introduce in  $T$ the coordinates $\xi^{\al}$,
the counterpart of the background inertial
coordinates $\bar\xi^{\al}$ in the space-time. 
By construction, the connection in the tangent space is zero
identically. For the connection in the space-time
in the the point~$P$ to be zero, too,  the coordinates are
to be related as 
$\xi^{\al}= \bar\xi^{\al} +{\cal O}((\bar\xi-\bar\Xi)^3)$.

Due to the spontaneous breaking, $GL(4,R)$ should be
realized in the nonlinear manner, with the
nonlinearity scale $M_A$, the Lorentz symmetry being still realized
linearly. The spinor representations of the latter correspond
conventionally to the matter fields. In this, the finite dimensional
spinors appear only at the level of $SO(1,3)$.
The broken part $GL(4,R)/SO(1,3)$ should be
realized in the Nambu-Goldstone mode. Accompanying  the spontaneous
emergence of the metric, there should appear the 10-component
Goldstone boson which corresponds to  the ten  generators of the
broken affine transformations. 
The  nonlinear realization  of the symmetry~$G$ spontaneously
broken to the symmetry $H\subset G$ can be built on the
quotient space $K=G/H$, the residual subgroup~$H$ serving as the
classification group. We are interested in the  
pattern $GL(4,R)/SO(1,3)$, with the quotient space consisting of all
the broken affine transformations.
Let $\kappa(\xi)\in K$ be the coset-function on the tangent space. To
restrict $\kappa$ by the quotient space, one should impose on the
representative group element some auxiliary condition, eliminating
explicitly the extra degrees of freedom.
Under the arbitrary affine transformation
$\xi \to  \xi'=A\xi+ a$, the coset is to transform~as
\be\label{eq: k}
(A,a)\ :\  \kappa(\xi) \to \ \kappa'(\xi')=
A \kappa(\xi)\Lambda^{-1},
\ee
where $\Lambda(\kappa,A)$ is the appropriate element of the residual
group, here the Lorentz one. This makes the transformed group element
compatible with the auxiliary condition.
In the same time, by the construction, the Minkowskian~$\eta$
is invariant under the nonlinear realization: 
\be\label{eq:eta}
(A,a)\ : \ 
\eta\to\eta'=\Lambda^{-1T}\eta\Lambda^{-1}=\eta
\ee
(in distinction with the linear representation
eq.~(\ref{eq:eta_lin})).

Otherwise, one can abandon any auxiliary condition
extending the affine symmetry by the hidden local symmetry 
$\hat H\simeq H$. In the tangent space, we should now distinguish  
two types  of indices: the Lorentz ones, acted on by the
local Lorentz transformations~$\Lambda(\xi)$, and the affine ones,
acted on by the global affine transformations~$A$. Designate the
Lorentz indices as $a, b$, etc,
while the affine ones as before $\al, \beta$, etc. 
The Lorentz indices are manipulated by
means of the Minkowskian $\eta_{ab}$ (respectively, $\eta^{ab}$).
The Goldstone field is represented by the arbitrary $4\times 4$
matrix ${\hat \kappa}^\al_a$ (respectively,  
${\hat \kappa}^{-1}{}^a_\al$) which  transforms
similar to eq.~(\ref{eq: k}) but with
arbitrary~$\Lambda(\xi)$. The extra 
Goldstone degrees of freedom are  unphysical due to the gauge 
transformations~$\Lambda(\xi)$. This is the linearization of
the nonlinear model, with the proper gauge boson 
being expressed, due to the equation on motion, through 
$\hat\kappa^a_\al$ and its derivatives. With this in mind, the abrupt
expressions entirely in terms of $\hat\kappa^a_\al$ and its
derivatives are used. The versions differ in the higher orders.

\paragraph{Matter and radiation}

Put for the matter fields $\phi$: 
\be\label{eq:phi}
\phi(\xi)\to\phi'(\xi')=\hat\rho_\phi(\Lambda)\phi(\xi),
\ee
with $\hat\rho_\phi$ taken in the proper Lorentz representations. 
As for the gauge bosons, they
constitutes one more separate kind of fields, the radiation. 
By definition, the gauge fields  $V_{\al}$ transform under
$A$ linearly as the derivative $\partial_{\al}=\partial/\partial
\xi^{\al}$. The modified fields
$\hat V_a = {\hat\kappa}^\al_a V_{\al}$
transform as the Lorentz vectors: 
\be\label{eq:barV}
\hat V(\xi)\to  \hat V'(\xi')=\Lambda^{-1T}  \hat V(\xi)
\ee
and are to be used in the model building.

\paragraph{Nonlinear model}

To explicitly account for the residual symmetry it is convenient to
start with the objects transforming only under the latter symmetry.
Clearly, any nontrivial combination of ${\hat\kappa}$
and ${\hat\kappa}^{-1}$ alone transforms explicitly under $A$. Thus
the derivative terms are inevitable. To this end, 
introduce the Cartan one-form:
\be\label{eq:Cartan}
{\hat\omega}=\eta {\hat\kappa}^{-1}d {\hat\kappa}.
\ee
The one-form transforms inhomogeneously under  the Lorentz group:
\be\label{eq:Delta_trans}
{\hat\omega}(\xi) \to
{\hat\omega}'(\xi')=\Lambda^{-1T}{\hat\omega}(\xi)\Lambda^{-1}
+ \Lambda^{-1T}\eta d\Lambda^{-1}.
\ee

By means of this one-form, one can define the nonlinear
derivatives of the matter fields $\hat D_a\phi$, the gauge strength
$\hat F_{ab}$, as well as the field strength for
affine Goldstone boson $\hat R_{abcd}$ and its contraction $\hat
R\equiv\eta^{ab} \eta^{cd}\hat R_{abcd}$.
The above objects can serve in turn 
as the building blocks for the  nonlinear model $GL(4,R)/SO(1,3)$ 
in the tangent space. Postulate the
equivalence principle in the sense that the  tangent space Lagrangian
should not depend explicitly on the background curvature
$\bar\rho^d{}_{abc}$ (cf.\ eq.~(\ref{eq:connection})).
Thus, the Lagrangian  may  be written as the general  Lorentz  (and,
thus, affine) invariant built of $\hat R$, 
$\hat F_{ab}$, $\hat D_a \phi$ and~$\phi$. As usually, one
restricts himself by the terms containing two derivatives at the most.

Such a Lagrangian being built, one can rewrite it by means of
${\hat\kappa}^\al_a$ and ${\hat\kappa}^{-1}{}^a_\al$ in the affine
terms. This clarifies the  geometrical
structure of the theory and  relates the latter with the gravity. 
The Lagrangian for the affine Goldstone boson,
radiation and matter becomes
\be\label{eq:L}
L= c_g M_A^2  R (\g_{\al\beta}) +
{L}_{r}(F_{\al\beta})
+  {L}_m (D_{\al}\phi,\phi ).
\ee 
Here 
\be\label{eq:gamma'}
\g_{\al\beta}={\hat\kappa}^{-1}{}^a_{\al}\eta_{ab}
{\hat\kappa}^{-1}{}^b_{\beta}
\ee
transforms as the affine tensor 
\be\label{eq:g_aff}
(A,a)\ : \  \g_{\al\beta}\to \g'_{\al\beta}=A^{-1}{}^\g{}_\al
\g_{\g\delta}A^{-1}{}^\delta{}_{\beta}.
\ee
It proves that $ R(\g_{\al\beta})=\hat R(\hat\omega)$ can
be expressed as the contraction $R=R^{\al\beta}{}_{\al\beta}$ of the
tensor $R^\g{}_{\al\delta\beta}\equiv \eta^{\g\g'}
{\hat\kappa}^{-1}{}^a_\al
{\hat\kappa}^{-1}{}^b_\beta   {\hat\kappa}^{-1}{}^d_\de
{\hat\kappa}^{-1}{}_{\g'}^c  \hat R_{cadb}$,
the latter in
turn being related with $\g_{\al\beta}$ as the Riemann-Christoffel
curvature tensor with the metric. In this, all the
contractions of the affine indices are understood with 
$\g_{\al\beta}$ (respectively, $\g^{\al\beta}$). 
Similarly,  $ D_{\al}\phi \equiv
{\hat\kappa}^{-1}{}_\al^a \hat D_a\phi$ looks like the  covariant
derivative of the  matter fields.
The gauge strength $F_{\al\beta}$ has the usual form containing the
partial derivative~$\partial_\al$.

\section*{General covariance}

\paragraph{General Relativity}

The preceding construction  referred to 
the tangent space~$T$ in the given point~$P$.  
Accept the so defined  Lagrangian  as that for the  space-time,
being valid in the background  inertial coordinates in the
infinitesimal neighbourhood of the point.
After  multiplying the Lagrangian by the
generally covariant volume element $(-\g)^{1/2}\,d^4
\bar\Xi$\,, with  $\g= \mbox{det}\g_{\al\beta}$, one gets
the infinitesimal contribution into the action in the given
coordinates.

The relation between the  background inertial coordinates and  the
world ones is achieved by means of the background  frame  
$\bar e^{\al}_\mu(X)$. In addition, introduce the effective frame
related with the background one as 
\be\label{eq:bare}
e_\mu^a(X)={\hat\kappa}^{-1}{}^a_{\al}(X)\,\bar e_\mu^{\al}(X).
\ee
The effective frame transforms  as  the Lorentz vector:
\be\label{eq:Le}
e_\mu(X)\to e'_\mu(X)=\Lambda(X)\, e_\mu(X).
\ee
Due to the local Lorentz transformations $\Lambda(X)$, one can 
eliminate six components out of~$e_\mu^a$, the latter having 
thus ten physical components. In this terms, the effective metric in
the world coordinates is
\be
g_{\mu\nu}\equiv  \bar e^\al_\mu \g_{\al\beta}\bar
e^\beta_\nu= e^a_\mu \eta_{ab}e^b_\nu.
\ee
In other words, the frame  $e^a_\mu$ defines the effective inertial
coordinates. Physically, eq.~(\ref{eq:bare}) describes the
disorientation of the effective inertial and background inertial
frames depending on the distribution of the affine Goldstone boson. 

By means  of $e^a_\mu$, the tangent space quantities transform in
the world coordinates to the usual
expressions of the Riemannian geometry containing  metric
$g_{\mu\nu}$ and the spin-connection $\omega_{ab\mu}$. One
gets for the total action:
\be\label{eq:L_AGB'}
I=\int \left(
-\frac{1}{2}M_{Pl}^2 R(g_{\mu\nu})
+\,L_r( F_{\mu\nu})
+L_m({D}_\mu \phi,\phi)\right)
(- g)^{1/2}\,d^4 X,
\ee
with $ g= \mbox{det}\, g_{\mu\nu}$. In the above, the constants
are adjusted so that $c_g M_A^2=1/2\,M^2_{Pl}\equiv 1/(16\pi
G_N)$, with $G_N$ being the Newton's constant. Thereof, one arrives at
the General Relativity (GR) equation for gravity.
Formally, the  Riemannian geometry  is valid
at any space-time intervals. Nevertheless, its accuracy  worsen
at the smaller intervals, requiring more  terms for
the decomposition in the ratio of energy to the symmetry
breaking scale~$M_A$,  the property characteristic  of
the effective theory. Thus,  the  Planck mass~$M_{Pl}\sim M_A$ is a
kind of the inverse minimal length in the nature.

\paragraph{Beyond the GR}

A priori, there are admissible the arbitrary Lagrangians in the
tangent space, satisfying the affine symmetry. The theory, generally
covariant in the tangent space, results in the theory generally
covariant in the space-time.
Under extension of the tangent space Lagrangians beyond the general
covariance,  the theory in the space-time ceases to be generally
covariant, too. The general covariance  moderates the
otherwise arbitrary  theories. Relative to  the
general coordinate transformations, the  GR extensions divide into
the inequivalent classes, each of which  is  characterised by the
particular set of the background parameter-functions.  A~priori,
no one of the sets is preferable. Which one is suitable (if any),
should be determined by observations. Each class consists of the
equivalent extensions related by the residual covariance group. 
Among the inequivalent extensions, there can be implemented the
natural hierarchy, according to whether the affine symmetry  is
explicitly  violated or not.

\paragraph{The Metauniverse}

Suppose that the origin  of the Universe lies in  the
actual transition between the two phases of the world continuum, the
affinely connected and metric ones. 
This transition is thus   the ``Grand Bang'', the origin of
the Universe in line with  the very space-time. 
There is conceivable the appearance, as well as 
disappearance and coalescence,  of the various regions of the metric
phase inside the affinely connected  one (or v.v.).
These metric regions are to be associated with the multiple universes,
constituting collectively the Metauniverse. Our Universe being one of
a lot, may clarify  the famous fine tuning problem.

\section*{Conclusion}

The concept of the metric space-time  as appearing in
the processes of the world structure formation, but not as a priori
given, is to change drastically the future comprehension of the
space-time, gravity and the Universe. There lies a long way ahead to
achieve this goal.

\end{document}